\documentclass[twocolumn,prl,amsmath,amssymb,superscriptaddress]{revtex4-2}
\usepackage{graphicx}% Include figure files
\usepackage{dcolumn}% Align table columns on decimal point
\usepackage{bm}% bold math
\usepackage{color}
\usepackage{amsmath, amssymb,amsfonts}
\usepackage{mathrsfs}
\usepackage{type1cm}
\usepackage[colorlinks,linkcolor=blue,citecolor=blue]{hyperref}
\usepackage{comment}
\usepackage{physics}

\begin{document}

\title{
Probing many-body localization crossover in quasiperiodic Floquet circuits on a quantum processor
}

\author{Kazuma Nagao}
\email{kazuma.nagao@riken.jp}
\affiliation{%
Computational Materials Science Research Team, RIKEN Center for Computational Science (R-CCS), Kobe, Hyogo 650-0047, Japan
}%
\affiliation{%
Quantum Computational Science Research Team, RIKEN Center for Quantum Computing (RQC), Wako, Saitama 351-0198, Japan
}%
%%%
\author{Tomonori Shirakawa}
\affiliation{%
Computational Materials Science Research Team, RIKEN Center for Computational Science (R-CCS), Kobe, Hyogo 650-0047, Japan
}%
\affiliation{%
Quantum Computational Science Research Team, RIKEN Center for Quantum Computing (RQC), Wako, Saitama 351-0198, Japan
}%
\affiliation{%
Quantum Mathematical Science Team, Division of Applied Mathematical Science, RIKEN Center for Interdisciplinary Theoretical and Mathematical Sciences (iTHEMS), Wako, Saitama 351-0198, Japan
}%
\affiliation{%
Quantum Computing Simulation Unit, Quantum-HPC Hybrid Platform Division, RIKEN Center for Computational Science, Kobe, Hyogo 650-0047, Japan
}%
\affiliation{%
Computational Condensed Matter Physics Laboratory, RIKEN Pioneering Research Institute (PRI), Wako, Saitama 351-0198, Japan
}%
%%%
\author{Rongyang Sun}
\affiliation{%
RIKEN Interdisciplinary Theoretical and Mathematical Sciences Program (iTHEMS), Wako, Saitama 351-0198, Japan
}%
\affiliation{%
Computational Materials Science Research Team, RIKEN Center for Computational Science (R-CCS), Kobe, Hyogo 650-0047, Japan
}%
%%%
\author{Peter Prelov\v{s}ek}
\affiliation{%
Jo\v{z}ef Stefan Institute, SI-1000 Ljubljana, Slovenia
}%
\affiliation{%
Faculty of Mathematics and Physics, University of Ljubljana, Slovenia.
}%
%%%
\author{Seiji Yunoki}
\affiliation{%
Computational Materials Science Research Team, RIKEN Center for Computational Science (R-CCS), Kobe, Hyogo 650-0047, Japan
}%
\affiliation{%
Quantum Computational Science Research Team, RIKEN Center for Quantum Computing (RQC), Wako, Saitama 351-0198, Japan
}%
\affiliation{%
Computational Condensed Matter Physics Laboratory, RIKEN Pioneering Research Institute (PRI), Wako, Saitama 351-0198, Japan
}%
\affiliation{%
Computational Quantum Matter Research Team, RIKEN Center for Emergent Matter Science (CEMS), Wako, Saitama 351-0198, Japan
}%

\date{\today}% It is always \today, today,

\begin{abstract}
Many-body localization (MBL) provides a mechanism by which interacting quantum systems evade thermalization, leading to persistent memory of initial conditions and slow entanglement growth. Probing these dynamical signatures in large systems and at long evolution times remains challenging for both classical simulations and current quantum devices.
Here we experimentally investigate the ergodic–MBL crossover in quasiperiodic Floquet Ising systems using up to 144 qubits on an IBM Quantum processor. By implementing deep Floquet circuits reaching up to 5000 cycles, we access long-time many-body dynamics beyond the regime explored in previous quantum computing experiments. Measurements of autocorrelation functions reveal a smooth crossover from rapid thermalization at weak quasiperiodic potential strength to persistent correlations in the strong-disorder regime. Notably, in addition to the one-dimensional system, we also observe clear signatures consistent with localization behavior in the two-dimensional system.
Furthermore, the quantum Fisher information exhibits logarithmic growth over thousands of Floquet cycles, providing evidence for slow entanglement spreading characteristic of the MBL regime.
These results demonstrate that programmable quantum processors can serve as experimental platforms for probing nonergodic quantum many-body dynamics and exploring localization phenomena in regimes beyond the reach of classical simulations.
\end{abstract}

\maketitle

Failure of thermal equilibration in isolated quantum many-body systems can emerge when local degrees of freedom are strongly coupled to a random or quasiperiodic potential~\cite{nandkishore2015many, altman2015universal, abanin2019colloquium, sierant2025many}. 
This phenomenon—known as many-body localization (MBL)—was originally discussed out of purely academic interest as a generalization of single-particle Anderson localization to interacting systems~\cite{basko2006metal, oganesyan2007localization}, 
before being realized experimentally in ultracold atomic systems~\cite{schreiber2015observation, choi2016exploring, luschen2017observation, lukin2019probing, leonard2023probing}. 
Exact diagonalization studies have provided evidence that
the ergodic regime of generic spin or electron systems in small clusters may continuously evolve into an MBL regime as the disorder strength increases~\cite{pal2010many, kjall2014many, luitz2015many, serbyn2015criterion, serbyn2016spectral}.
Moreover, finite-size MBL is not restricted to genuinely isolated systems but also arises in periodically driven Floquet systems~\cite{abanin2019colloquium, ponte2015periodically, lazarides2015fate, zhang2016a, sierant2023stability, falcao2024many}.
Floquet systems are particularly interesting in this context,
as periodic driving generically leads to heating toward an infinite temperature state 
unless mechanisms such as localization prevent thermalization~\cite{else2020discrete, mi2022time}.
Despite extensive studies, it remains under active debate 
whether the MBL regime can be sharply distinguished from the ergodic regime even in the thermodynamic limit~\cite{thiery2018many, morningstar2022avalanches}.

Deep in the MBL regime,
even highly-excited eigenstates are expected to exhibit area-law entanglement scaling~\cite{abanin2019colloquium, bauer2013area}. 
Tensor-network methods therefore provide a suitable approach to many-body systems dominated by such low-entangled states~\cite{schollwock2011density, cirac2021matrix}. 
Studies on one-dimensional MBL systems using matrix product states (MPS) have quantitatively demonstrated characteristic features of deep MBL, 
such as logarithmically slow spreading of entanglement, even for system sizes beyond the capability of exact diagonalization~\cite{znidaric2008many, bardarson2012unbounded, khemani2016obtaining, yu2017finding, chanda2020time}. 
Nevertheless, tensor-network simulations in the crossover regime with intermediate disorder strength become challenging,
as the coexistence of localized and thermal eigenstates
leads to substantial entanglement growth and a rapid increase of the required bond dimension.
Moreover, it remains highly nontrivial whether properties observed in such low entangled one-dimensional systems extend to higher spatial dimensions~\cite{chandran2016many, potirniche2019exploration, wahl2019signatures, doggen2020slow}.
In general, higher-dimensional tensor-network states such as projected entangled pair states (PEPS) require approximate tensor contraction schemes, since even at fixed bond dimension the exact contraction of PEPS typically entails an exponential growth of computational cost with system size~\cite{schuch2007computational}. Consequently, practical simulations rely on approximate algorithms and remain computationally demanding even with moderate bond dimensions~\cite{kshetrimayum2020time, dziarmaga2022simulation}, which limits our understanding of the stability of MBL against dimensionality.
These limitations highlight the need for alternative computational paradigms capable of probing complex many-body phenomena that are inaccessible to classical simulations.

Recent advances in quantum processors have opened utility-scale applications to computationally challenging many-body problems~\cite{kim2023evidence, robledo2025chemistry, shinjo2024unveiling_p, switzer2026realization},
including MBL~\cite{smith2019simulating, shtanko2025uncovering, hayata2025digital, lunkin2026evidence}. 
Hardware platforms implementing sets of one- and two-qubit gates provide an ideal setup for digital quantum simulations of Floquet systems. 
On current noisy processors, it is crucial to suppress the influence of various errors and the accumulation of circuit infidelities, and to recover signals from readouts using error mitigation techniques~\cite{cai2023quantum, aharonov2025reliable}.
Exploring strategies that enable sufficiently deep cycles of a Floquet circuit would represent a substantial step toward systematically examining nonergodicity across the localization-delocalization crossover.

Here, we report the observation of the ergodic–MBL crossover in quasiperiodic (QP) Floquet Ising systems using an IBM Quantum Heron processor, {\tt ibm\_kobe}. The experiments are performed on both a 129-qubit chain and a 144-qubit heavy-hexagonal lattice, enabling the investigation of localization dynamics in one- and two-dimensional geometries. By implementing deep Floquet circuits reaching up to 5000 cycles, we access long-time many-body dynamics well beyond the regime explored in previous quantum computing experiments.
A key ingredient enabling such long-time evolution is the use of hardware-native fractional gates available in the latest Heron processors~\cite{ibm_document_migrate_fractional_gate}. These gates allow direct implementation of RZZ two-qubit interactions with continuous angles, substantially reducing circuit depth compared with conventional decompositions into Clifford gate sets. 
In addition, all transmon qubits are initialized in their ground state, which suppresses decay processes associated with excited-state relaxation during circuit evolution. 

Measurements of autocorrelation functions reveal a smooth crossover from rapid thermalization at weak quasiperiodic potential strength to persistent correlations in the strong-disorder regime. Notably, we observe clear signatures consistent with localization behavior not only in the one-dimensional system but also in the two-dimensional geometry. Furthermore, the quantum Fisher information, which serves as a proxy for entanglement entropy, exhibits logarithmic growth over thousands of Floquet cycles, providing evidence for slow entanglement spreading characteristic of the MBL regime. 
The experimental results are further supported by tensor-network simulations for up to 100 cycles and by state-vector simulations of a 28-qubit system for up to 5000 cycles.

%%%%%%%%%%%%%%%%%%%%
%%%%%%%%%%%%%%%%%%%%
\begin{figure*}
\includegraphics[width=\textwidth]{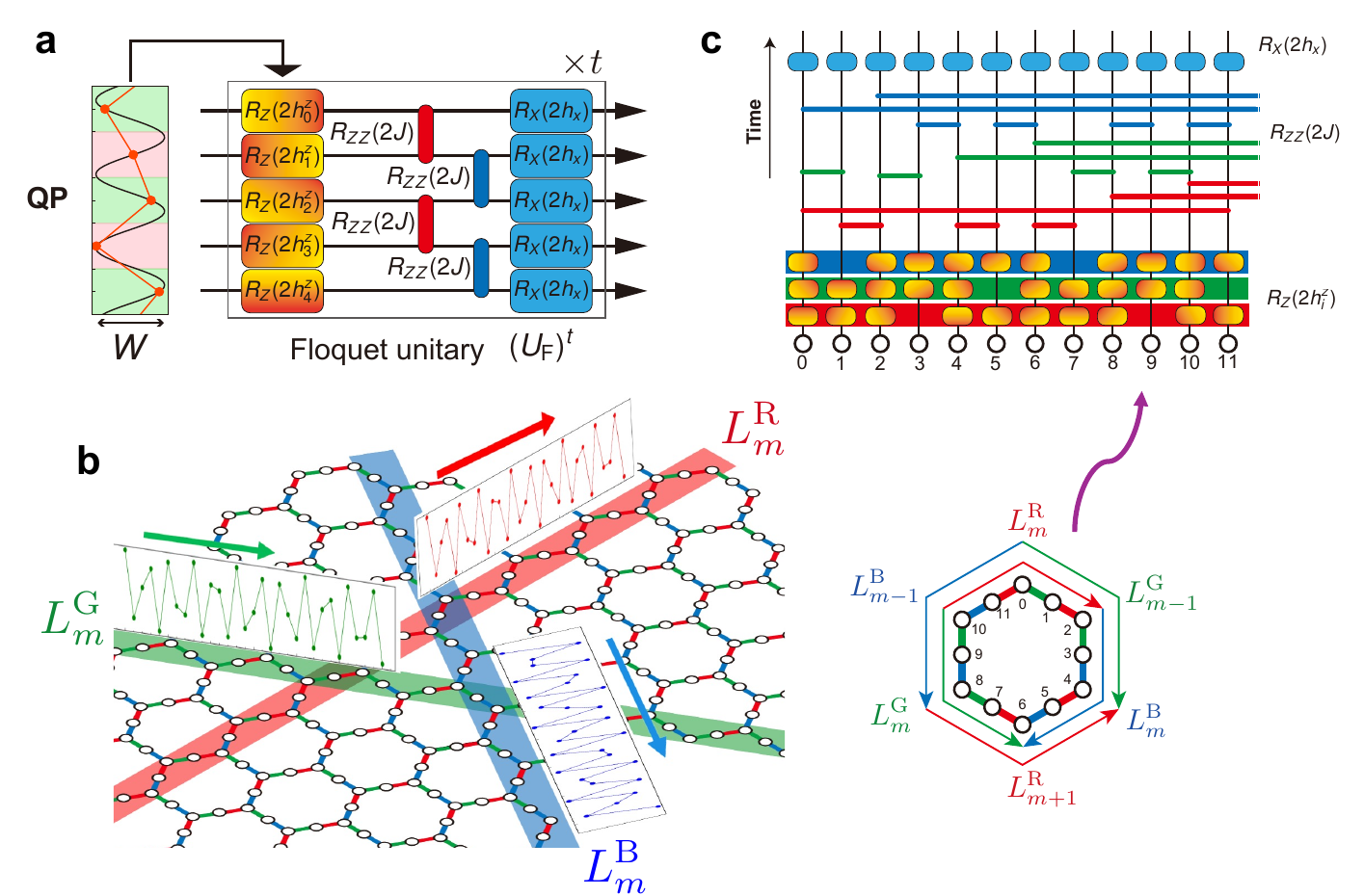}
\vspace{0mm}
\caption{
{\bf 
Kicked Ising systems subjected to quasiperiodic (QP) potentials. 
}
{\bf a} 
Circuit representation of the one-dimensional kicked Ising Floquet operator ${\hat U}_{\rm F}^{(1)}$. 
A QP potential with strength $W$ is applied through the layer of RZ gates, which induces localization in the Hilbert space of qubits. 
The Ising coupling is decomposed in two layers of RZZ gates (red and blue),
followed by a global RX rotation. 
The boxed block indicates a single Floquet cycle, which is repeated $t$ times. 
{\bf b}~Schematic illustration of the two-dimensional kicked Ising model with QP potentials. 
The Floquet circuit acts on a heavy-hexagonal lattice and includes red, blue, and green coupling layers. 
A QP potential is applied along $L^{\rm R/G/B}_{m}$ ($m = 0, 1, 2, \cdots$), indicated by colored stripes (red, green, or blue) passing through the qubits and forming one-dimensional paths. 
Each hexagon in the lattice therefore contains three different stripe directions.  
{\bf c}~Part of the two-dimensional Floquet circuit ${\hat U}_{\rm F}^{(2)}$ acting on a single hexagonal unit. 
The RZ and RZZ gates are arranged in three colored layers. 
}
\label{fig1}
\end{figure*}
%%%%%%%%%%%%%%%%%%%%
%%%%%%%%%%%%%%%%%%%% 

We consider two types of periodically driven spin-1/2 systems. 
The first model is a one-dimensional kicked Ising system governed  
by the Floquet unitary operator~\cite{bertini2018exact}
\begin{align} \label{eq: floquet_unitary_1d}
{\hat U}_{\rm F}^{(1)} = {\hat U}_{X} e^{- {\rm i} J \sum_{j=0}^{N-2} {\hat Z}_{j} {\hat Z}_{j+1} - {\rm i} \sum_{j=0}^{N-1} h^z_{j} {\hat Z}_{j}},
\end{align}
where $j$ are indices of qubits, and 
the global $X$-rotation operator is defined as 
\begin{align}
{\hat U}_{X} =  e^{- {\rm i} h_x \sum_{j=0}^{N-1}{\hat X}_{j}}.
\end{align}
In Eq.~(\ref{eq: floquet_unitary_1d}), 
the longitudinal magnetic fields $h^z_j$ are spatially modulated as an incommensurate cosine potential relative to the lattice spacing, i.e., 
\begin{align}
h^z_j = \frac{W}{2} \cos (2 \pi \beta j + \omega_0),
\end{align}
where $\beta = (\sqrt{5}-1)/2$ is the inverse golden ratio~\cite{falcao2024many}, $W$ is the potential strength, and $\omega_0$ is a phase offset. 
In this work, we set $\omega_0=0$.
A time evolved state after $t$ cycles is $\ket{\psi_t} = ({\hat U}^{(1)}_{\rm F})^t \ket{\psi_0}$, 
where $\ket{\psi_0}$ is an initial state. 
Equation (\ref{eq: floquet_unitary_1d}) has a circuit representation with local RX, RZ, and RZZ gates as  
\begin{align}
{\hat U}_{\rm F}^{(1)}  = \prod_{j=0}^{N-1}{\hat R}_{X}^{(j)}(2h_x) \prod_{j=0}^{N-2}{\hat R}_{ZZ}^{(j,j+1)}(2J) \prod_{j=0}^{N-1}{\hat R}_{Z}^{(j)}(2h_j^{z}),
\end{align}
where the middle RZZ coupling layer can be exactly decomposed into two independent Trotter layers in one dimension (Fig.~\ref{fig1}a).

Throughout this work,
we set $J=h_x=W^{-1}$ and hence weak (strong) potential strength $W$ corresponds to the ergodic (MBL) regime and to strong (weak) effective coupling
\footnote{
In general, 
the ergodic-MBL crossover does not rely on this specific parametrization.
For instance, 
choosing $W = 2\pi$ with $J = h_x = w^{-1}$ ($w \in {\mathbb R}_{>0}$) 
yields an alternative setting that reproduces similar crossover behavior~\cite{sierant2023stability}.
}.
For this parametrization, 
$W = 4/\pi$ marks the solvable self-dual point, at which the exact solution of the spectral form factor exhibits maximally chaotic behavior, implying the absence of MBL~\cite{bertini2018exact}.

The second model is a two-dimensional kicked Ising system that comprises three coupling layers on a heavy-hexagonal lattice, where QP potentials are applied along three different lattice directions (Fig.~\ref{fig1}b).
The Floquet unitary operator is given by 
\begin{align} \label{eq: floquet_unitary_2d}
{\hat U}_{\rm F}^{(2)} = {\hat U}_{X} ({\hat U}^{\rm B}_{ZZ}{\hat U}^{\rm G}_{ZZ}{\hat U}^{\rm R}_{ZZ}) ({\hat U}^{\rm B}_{Z}{\hat U}^{\rm G}_{Z}{\hat U}^{\rm R}_{Z}),
\end{align}
where
each of the sub-cycles ${\hat U}^{\alpha}_{ZZ}$ and ${\hat U}^{\alpha}_{Z}$ belongs to a colored layer ($\alpha \in \{ {\rm R}, {\rm G}, {\rm B} \}$) 
and 
is defined as 
\begin{align}
{\hat U}^{\alpha}_{ZZ} 
&= \prod_{(j,l) \in \Lambda_{\alpha}} {\hat R}^{(j,l)}_{ZZ}(2J), \\
{\hat U}^{\alpha}_{Z} 
&= \prod_{m=0}^{M_{\alpha}-1} \prod_{j \in L^{\alpha}_{m}} {\hat R}^{(j)}_{Z}[2 h^{z}_{j}(W_\alpha)].
\end{align}
Here, $L^{\alpha}_{m}$ denotes parallel stripes of qubits along one of the three lattice directions, as illustrated in Fig.~\ref{fig1}b, 
The index $m$ labels the stripes forming one-dimensional paths along direction $\alpha$, 
and $M_{\alpha}$ is the number of parallel $\alpha$ stripes covering the lattice. 
Moreover, each set $\Lambda_{\alpha}$ defines the nearest-neighbor Ising couplings and consists of pairs of qubits connected by $\alpha$-colored bonds.
The schematic structure of Eq.~(\ref{eq: floquet_unitary_2d}) is illustrated in Fig.~\ref{fig1}c, where 
$\Lambda_{\rm R} = [(1,2), (4,5), (6,7), (0,11), \cdots]$,
$\Lambda_{\rm G} = [(0,1), (2,3), (7,8), (9,10), \cdots]$,
$\Lambda_{\rm B} = [(3,4), (5,6), (8,9), (10,11), \cdots]$,
and $\Lambda_{\rm R} \cap \Lambda_{\rm G} \cap \Lambda_{\rm B} = \emptyset$.
In what follows, we assume equal-strength QP potentials with $W_{\alpha} = W$ for all $\alpha$, 
and set $J=h_x = W^{-1}$,
as in the one-dimensional case.

%%%%%%%%%%%%%%%%%%
\section{Results}
%%%%%%%%%%%%%%%%%%

%%
{\bf Autocorrelation function.}
Broken ergodicity in MBL systems
can be probed through an averaged autocorrelation function~\cite{switzer2026realization}, 
defined as 
\begin{align}
   A(t) 
   &= \frac{1}{N}\sum_{i=0}^{N-1} \bra{\psi_0} {\hat Z}_{i}(t) {\hat Z}_{i} \ket{\psi_0} \\ 
   &= \frac{1}{N} \sum_{i=0}^{N-1} (1-2\sigma_i) \bra{\psi_t} {\hat Z}_{i} \ket{\psi_t}, \nonumber 
\end{align}
where 
$\ket{\psi_0} = \otimes_{j}\ket{\sigma_j}$ ($\sigma_i \in \{0,1\}$) is a direct product state in the computational Pauli-$Z$ basis,
and thus the onsite time correlation reduces to local magnetization.  
Moreover, 
we select the ferromagnetic all-up state $\ket{\psi_0} = \ket{0} \otimes \cdots \otimes \ket{0}$ as the initial state,
which is an array of ground state transmon qubits. 
This choice is essential to maintain the initial magnetization value even after sufficiently deep cycles, which is crucial for addressing longtime nonergodic dynamics. 
We find that any other initial state leads to unavoidable attenuation of the measured magnetization, 
predominantly caused by natural decay of excited qubits in the initial state, 
see Supplementary Information.

%%%%%%%%%%%%%%%%%%%%
%%%%%%%%%%%%%%%%%%%%
\begin{figure*}
\includegraphics[width=\textwidth]{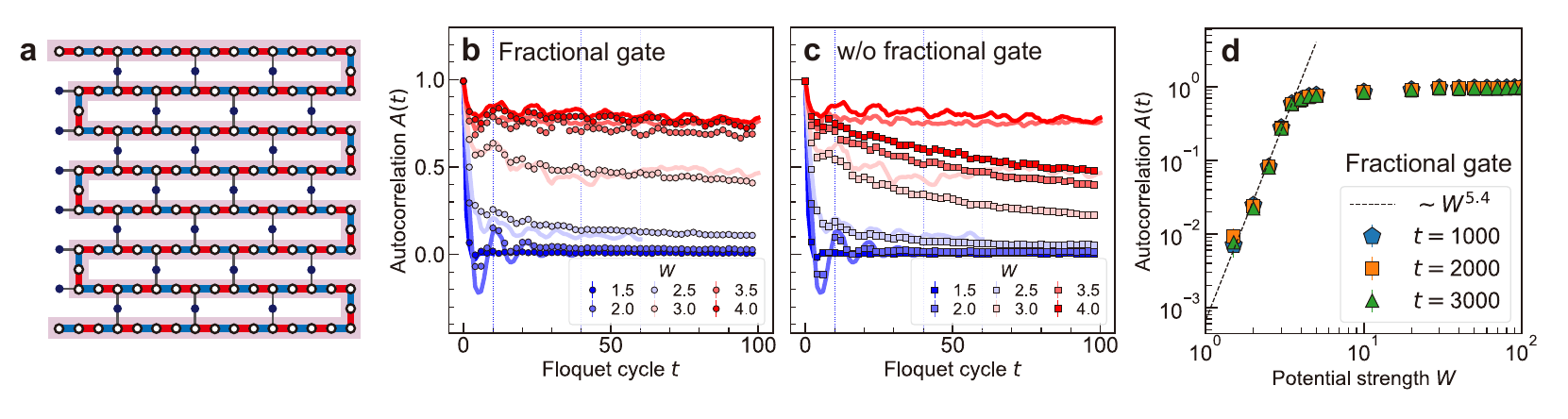}
\vspace{-6mm}
\caption{
{\bf 
Autocorrelation function in the one-dimensional quasiperiodic (QP) kicked Ising model. 
}
{\bf a} 
Connectivity layout of qubits on the {\tt ibm\_kobe} processor.
White circles highlight the 129 qubits selected for our experiments, forming a one-dimensional chain. 
Colored bonds represent nearest-neighbour RZZ couplings.
Red (blue) bonds indicate the even (odd) layer of parallelized RZZ gates (see also Fig.~\ref{fig1}a).
{\bf b} 
Experimental results obtained using fractional gates up to 100 Floquet cycles. 
Symbol colors indicate the strength of $W$. 
Error bars are estimated from the average measurement errors of ${\hat Z}_j$, but are sufficiently small to be negligible. 
For comparison, tDMRG results are shown as solid lines with the corresponding colors, up to the Floquet cycle $t$ for which convergence is confirmed with bond dimension $\chi=512$ (indicated by vertical dotted lines).
{\bf c} 
Same as {\bf b} but without fractional gates. 
{\bf d} 
Longer-time evolution of the autocorrelation up to $3000$ Floquet cycles using fractional gates. 
The dashed line shows a power-law fit to the data with $W < 4.0$. 
}
\label{fig2}
\end{figure*}
%%%%%%%%%%%%%%%%%%%%
%%%%%%%%%%%%%%%%%%%%

Experimental results of the autocorrelation function for the one-dimensional kicked Ising model are shown in Fig.~\ref{fig2}. 
In our experiments, we used the {\tt ibm\_kobe} processor and executed the one-dimensional Floquet circuits with 129 qubits, whose qubit connectivity is presented in Fig.~\ref{fig2}a. 
The RX, RZ, and RZZ gates in circuits are directly implemented on hardware using fractional gates. 
Figure~\ref{fig2}b demonstrates the effectiveness of fractional gates, as our experimental results reasonably reproduce the tensor-network results based on the time-dependent density matrix renormalization group (tDMRG)~\cite{sun2024improved}.  
For $W \lesssim 3$, 
the measured autocorrelation function exhibits rapid decay, 
implying Floquet thermalization to an infinite temperature state of the qubit system.
The decay process becomes progressively slower with $W$, 
signaling failure of thermalization. 
Notably, the experimental results capture the ergodic behavior beyond the reach of classical simulations, 
where tDMRG fails to converge up to the maximum bond dimension ($\chi = 512$).

The advantage of fractional gates becomes more pronounced when compared with circuits transpiled to the native Clifford gate set, which typically result in increased circuit depth (see Methods).
Figure~\ref{fig2}c shows experimental results for the transpiled Floquet circuits, in which each fractional RX or RZZ gate of ${\hat U}^{(1)}_{\rm F}$ is decomposed into the native gate set consisting of CZ, X, SX and RZ.
In contrast to the tDMRG results, the measured correlation decays even for $W \gtrsim 3.0$.
The breakdown of the MBL behavior can be attributed to accumulated errors from SX and CZ gates, whose interplay reduces the fidelity to the time-evolved state (see Supplementary Information).

To further validate the measured nonergodic behavior, we probe the evolution at much longer times up to $t = 3000$ using fractional gates.
Figure~\ref{fig2}d plots the autocorrelation function for $t \ge 1000$ as a function of the potential strength $W$.
The localization behavior persists even at such long time steps, indicating the stability of localization in the experiment.
For $W \lesssim 4.0$, the autocorrelation can be empirically fitted by a power-law dependence, yielding $A(t \gg 1) \sim W^{5.4}$.
In the limit $W \rightarrow \infty$, the autocorrelation approaches unity, corresponding to the bond-decoupling limit in which the system decomposes into independent sites and no entanglement is formed.

%%%%%%%%%%%%%%%%%%%%
%%%%%%%%%%%%%%%%%%%%
\begin{figure*}
\includegraphics[width=\textwidth]{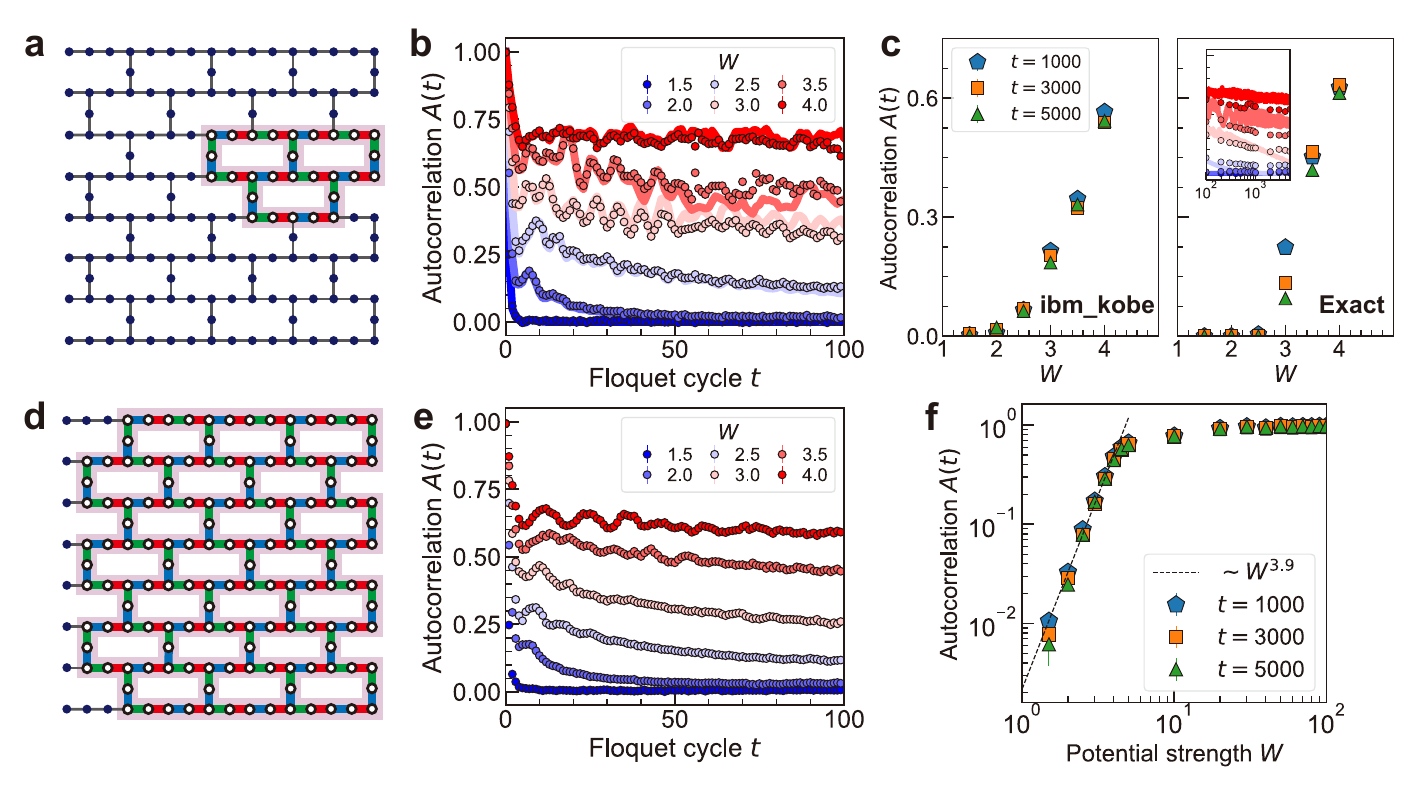}
\vspace{-6mm}
\caption{
{\bf 
Autocorrelation function in the two-dimensional quasiperiodic (QP) kicked Ising model. 
}
{\bf a} 
Connectivity layout for the 28-qubit system in two dimensions. 
{\bf b} 
Measured autocorrelation function $A(t)$ for the 28-qubit system for $t \leq 100$. 
Symbol colors indicate different values of $W$.
Solid lines show the corresponding state-vector simulation results. 
{\bf c} 
Left: long-time evolution of the 28-qubit system up to 5000 Floquet cycles.
Right: corresponding state-vector simulation results. 
The inset extends the experimental data in {\bf b} to longer times up to 5000 Floquet cycles.
Symbol colors correspond to those in {\bf b}.
{\bf d} 
Same as {\bf a} but for the 144-qubit system. 
{\bf e} 
Same as {\bf b} but for the 144-qubit system.
{\bf f} 
Long-time evolution of the 144-qubit system up to 5000 Floquet cycles. 
The dashed line shows a power-law fit to the data with $W < 4$.
}
\label{fig3}
\end{figure*}
%%%%%%%%%%%%%%%%%%%%
%%%%%%%%%%%%%%%%%%%%

Fractional gates enable us to observe similar localization even in the two-dimensional kicked Ising model.
To study this, we begin by comparing the experimental results for 28 qubits with state-vector simulations.
The geometry on hardware is shown in Fig.~\ref{fig3}a.
Figure~\ref{fig3}b presents the measured autocorrelation function up to 100 cycles,
which is in good agreement with the state-vector simulations.
Deeper cycle evolution for $t \gg 100$ up to 5000 cycles is only captured qualitatively (inset in Fig.~\ref{fig3}c), while the measured $W$ dependence is similar to the numerically exact one (Fig.~\ref{fig3}c). 
In the experimental results, the localization signature is visible in sharp for $W \gtrsim 3.0$ and the finite correlations are almost converged for sufficiently deep cycles.
However, we find that even for $W \lesssim 2.5$, which should correspond to the ergodic regime according to the exact results, non-negligible correlations remain. 
This may be attributed to accumulated errors during the circuit evolution.
As discussed below, this behavior cannot be explained solely by a simple noise model, implying the necessity of more sophisticated error mitigation and suppression techniques beyond the scope of this work.

Next, we consider a larger two-dimensional system with 144 qubits, 
which is classically intractable for exact state-vector simulations and beyond the practical reach of tensor-network approaches such as MPS for long-time dynamics.
The qubit geometry is shown in Fig.~\ref{fig3}d.
The Floquet circuit contains 54 red, 55 blue, and 55 green bonds.
Experimental results for up to 100 Floquet cycles are shown in Fig.~\ref{fig3}e. 
Similar to the 28-qubit case, the autocorrelation function exhibits localization signatures for $W \gtrsim 3.0$.
The slow relaxation behavior persists up to 5000 Floquet cycles, as shown in Fig.~\ref{fig3}f.
As in the one-dimensional results in Fig.~\ref{fig2}d, a power-law fit yields $A(t \gg 1) \sim W^{3.9}$ for $W \lesssim 4.0$, 
while the autocorrelation approaches unity in the large $W$ limit.

{\bf Entanglement spreading.}
A characteristic feature expected in the MBL regime is how locally encoded quantum information spreads under unitary dynamics, 
leading to logarithmically slow growth of entanglement from an initially unentangled state. 
In principle, this behavior can be detected by measuring the subsystem von Neumann entanglement entropy (see Supplementary Information). 
However, such measurements are difficult to implement in experiments. 
Instead, we evaluate a quantum Fisher information (QFI) for the fully polarized initial state~\cite{smith2016many, guo2021observation},
defined as the sum of all-to-all correlation functions of Pauli-$Z$ operators 
\begin{align} \label{eq: qfi_ferro}
F_{Q} = 4 \sum_{i,j }\left( \langle {\hat Z}_i {\hat Z}_j \rangle -  \langle {\hat Z}_i \rangle \langle {\hat Z}_j \rangle \right).
\end{align}
This quantity serves as an entanglement witness that captures deviations from an unentangled state. 
The QFI in the form of Eq.~(\ref{eq: qfi_ferro}) can be readily reconstructed from measured bitstring samples obtained in shots (see Methods).

%%%%%%%%%%%%%%%%%%%%
%%%%%%%%%%%%%%%%%%%%
\begin{figure*}
\includegraphics[width=\textwidth]{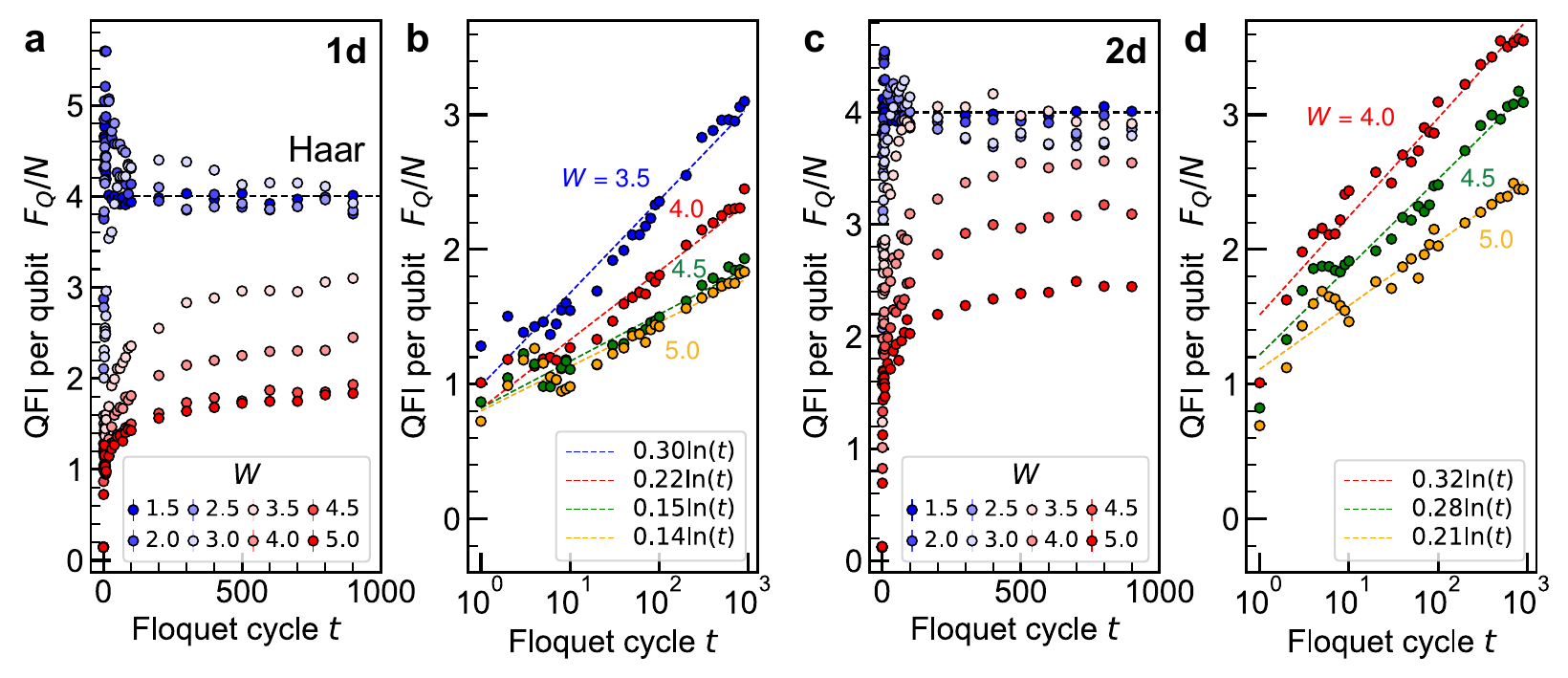}
\vspace{-6mm}
\caption{
{\bf Time evolution of the quantum Fisher information (QFI).}
{\bf a}
Measured QFI for the one-dimensional kicked Ising model with 129 qubits.
For $W \lesssim 3.0$, the QFI rapidly approaches the Haar random value $F_{Q}^{\rm Haar} \approx 4/N$, indicated by the dashed black line. 
For $W \gtrsim 3.5$, the spreading of correlations is suppressed, 
leading to logarithmic growth characteristic to the MBL regime. 
The QFI is evaluated from 
$2^{14}=16384$ measurement shots.
Error bars are obtained using the bootstrap method applied to the measured bitstrings. 
{\bf b}
Numerical fit of the one-dimensional results in the MBL regime. 
The data are well described by a logarithmic function $a + b \ln t$. 
{\bf c}
Same as {\bf a} but for the two-dimensional kicked Ising model with 144 qubits. 
{\bf d}
Same as {\bf b} but for the two-dimensional case. 
Logarithmic growth is observed for $W \gtrsim 4$, whereas for smaller $W$ the QFI approaches the Haar random value.  
}
\label{fig4}
\end{figure*}
%%%%%%%%%%%%%%%%%%%%
%%%%%%%%%%%%%%%%%%%%

Figure~\ref{fig4}a shows experimental results for the QFI in the one-dimensional kicked Ising model on the 129-qubit geometry. 
For $W \lesssim 3.0$, the QFI rapidly approaches the thermal value $F_{Q}^{\rm Haar} \approx 4/N$,
signaling thermalization to an infinite-temperature state described by Haar-random states~\cite{page1993average, shtanko2025uncovering}. 
For $W \gtrsim 3.5$, the growth of the QFI is suppressed and it does not reach the thermal value even after 1000 cycles. 
In this regime, as shown in Fig.~\ref{fig4}b, the QFI grows logarithmically, consistent with the MBL regime. 
Similar logarithmic behavior is observed in state-vector simulations up to 16 qubits (see Supplementary Information). 
Comparing with the autocorrelation results above, 
we identify an intermediate regime for $2.0 \lesssim W \lesssim 3.0$ in the 129-qubit system. 
In this regime, the autocorrelation function retains a finite residue while the QFI is already close to the thermal value. 
This mixed behavior indicates a crossover between ergodic and nonergodic dynamics.
Hence, the threshold for the MBL regime is estimated as $W_{*} \sim 3.5$.
This estimation is found to be reasonably consistent with the state-vector simulations for the 16-qubit system.

Figure~\ref{fig4}c shows the QFI results for the two-dimensional kicked Ising model on the 144-qubit geometry. 
For $W \lesssim 3.5$, the QFI grows rapidly and approaches the thermal (Haar-random) value for $t \gtrsim 100$. 
For $W \gtrsim 4.0$, the QFI instead exhibits logarithmic growth, as shown in Fig.~\ref{fig4}d.
The intermediate region $2.5 \lesssim W \lesssim 3.5$ corresponds to a crossover regime in which the autocorrelation retains a finite long-time value. 
Consequently, the apparent threshold for the MBL regime shifts to larger $W$ in two dimensions, with $W_{*} \sim 4$.
Furthermore, in contrast to the one-dimensional case, the QFI per qubit tends to take larger values, possibly reflecting the higher connectivity of the heavy-hexagonal lattice.

%%%%%%%%%%%%%%%%%%
\section{Discussion}
%%%%%%%%%%%%%%%%%%

We have presented a comprehensive investigation of the ergodic–MBL crossover in quasiperiodic kicked Ising systems using a state-of-the-art quantum processor. Our results demonstrate that many-body Floquet dynamics involving more than 100 qubits and extending to $O(10^3)$ Floquet cycles can be accessed on current noisy quantum computers when physically stable initial states are employed. These capabilities open a new avenue for exploring outstanding problems in many-body localization, such as quantum avalanches~\cite{thiery2018many, morningstar2022avalanches}, and illustrate the potential of programmable quantum processors as experimental platforms for nonequilibrium quantum many-body physics. Future experiments combining fractional gates with advanced error suppression and mitigation techniques represent a promising direction. 

A striking feature of our results is the robustness of the observed localization signatures over long evolution times. The choice of the initial condition plays a crucial role in achieving such stability, as it minimizes the population of excited transmon states and thereby suppresses decay processes associated with their finite lifetime. In this setting, the use of fractional gates enables the Floquet circuit to be implemented with substantially reduced circuit depth, without requiring tailored error-mitigation protocols such as zero-noise extrapolation. 

To better understand the role of hardware noise, we analyzed the experimental data using a phenomenological depolarizing noise-channel model. The observed dynamics cannot be fully explained by this simple stochastic noise model (see also Supplementary Information). This discrepancy suggests that, in the fractional-gate mode, biased coherent errors may play a more significant role than incoherent errors described by stochastic noise channels. 
These findings highlight the potential of quantum processors to investigate long-time nonequilibrium dynamics of quantum many-body systems in regimes that remain challenging for classical simulations.

%%%%%%%%%%%%%%%%%%
\section{Methods}
%%%%%%%%%%%%%%%%%%

{\bf Fractional gates.}
The fractional RZZ and RX gates referred to here are continuous-angle operations included in the native gate set of the 156-qubit Heron processors~\cite{abughanem2024ibm, ibm_document_migrate_fractional_gate}. 
The rotation angle $\theta_{zz}$ of RZZ gate is restricted in the finite range $0 < \theta_{zz} \leq \pi/2$, 
thereby leading to the allowed range of $W$ as $4/\pi \leq W < \infty$ in $J = \theta_{zz}/2 = 1/W$. 
The angle $\theta_{x}$ of RX gate is arbitrary. 
In general, 
each fractional gate is non-Clifford, unless the angle takes Clifford values as $\theta_{zz} = p \pi/2$ or $\theta_x = p \pi /2$, where $p$ is an integer.
The full set of native gates is given by 
\begin{align}
\{R_X(\theta_x), R_Z(\theta_z), R_{ZZ}(\theta_{zz}), {\rm CZ}, {\rm Id},  {\rm SX}, X\},
\end{align}
where ${\rm CZ}, {\rm Id},  {\rm SX}, X$ are Clifford gates. 
The RZ gate can also be considered as a fractional gate, but it is trivial as it can be implemented without shining microwave pulses.

A quantum circuit that only consists of fractional RZZ gates and therefore commutes with Pauli-$Z$ operators
facilitates insights 
on the effectiveness of fractional gates. 
If we execute this circuit on a quantum processor with the fully polarized initial state $\ket{0\cdots 0}$, 
the expectation value of the total magnetization 
$\langle {\hat M}_z \rangle = N^{-1}\sum_{j} \langle {\hat Z}_j \rangle$
appears to be preserved during cycles and stay almost at unity. 
By contrast, 
replacing each RZZ gate with a transpiled gate consisting of two CZ gates and four SX gates,
the magnetization shows exponential decay with the circuit depth, 
since the SX gates are off-diagonal (see Supplementary Information).

{\bf Experimental details.}
The experimental data in Figs.~\ref{fig2} and~\ref{fig3} were obtained using the {\tt Estimator V2} primitive in {\tt Qiskit Runtime}. 
The resilience level was set to one,
which activates only light error mitigation associated with measurement and readout. 
In the transpilation step, we explicitly disabled automatic circuit optimization to avoid undesired gate fusion and to preserve the circuit structure. 
The estimation precision of expectation values was $\epsilon_{\rm p} = 2^{-6} \approx 0.016$.
Furthermore, the data in Fig.~\ref{fig4} were obtained using the {\tt Qiskit Sampler V2}.
The number of shots was $2^{14} = 16384$ for each circuit. 
Typical specifications of the {\tt ibm\_kobe} processor at the time of execution are summarized in Supplementary Information.

In our experiments, we prepared a set of circuits $\{ {\cal C}_0, {\cal C}_1, \cdots, {\cal C}_t, \cdots \}$, where ${\cal C}_t$ represents a Floquet circuit with $t$ cycles, 
and submitted them simultaneously to the {\tt ibm\_kobe} processor.
Each block of the circuit corresponds to the Floquet unitary operator ${\hat U}^{(1)}_{\rm F}$ or ${\hat U}^{(2)}_{\rm F}$, 
and consists of a layer of RZ gates, multiple layers of RZZ gates, and a layer of RX gates. 
Between these layers, we insert barriers across all qubits, 
ensuring that fractional gates within each layer are executed simultaneously on the hardware, and thereby minimizing qubit idle time.
Executing the circuit ${\cal C}_t$ yields the time-evolved state $\ket{\psi_t} = {\cal C}_{t}\ket{\psi_0}$, followed by measurements of ${\hat Z}_{j}$. 
For the one-dimensional (two-dimensional) kicked Ising model with 129 (144) qubits, each Floquet cycle contains 128 (164) RZZ gates arranged in 4 (7) circuit layers, including the RX and RZ layers.

%\bibliography{ref}

{\bf Acknowledgments}

%%%%%%%%%%%%%
\begin{acknowledgments}
%%%%%%%%%%%%%

We thank Mirko Amico, 
Ori Alberton, 
Kazuya Shinjo, 
Niall Robertson, 
Netanel Lindner, 
Kazuhiro Seki, 
and Antonio Mezzacapo for valuable comments and discussions.
This work was partially supported by Project No.~JPNP20017, funded by the New Energy and Industrial Technology Development Organization (NEDO), Japan. 
We acknowledge support from JSPS KAKENHI (Grant Nos.~JP21H04446 and JP25K17321)
from the Ministry of Education, Culture, Sports, Science and Technology (MEXT), Japan.
Additional funding was provided by 
JST COI-NEXT (Grant No.~JPMJPF2221) and by 
the Program for Promoting Research of the Supercomputer Fugaku (Grant No.~MXP1020230411) from MEXT, Japan.  
We further acknowledge support from 
the RIKEN TRIP initiative (RIKEN Quantum), the UTokyo Quantum Initiative, and
the COE research grant in computational science from Hyogo Prefecture and Kobe City through the Foundation for Computational Science. 
MPS simulations are based on the high-performance tensor computing library {\it GraceQ/tensor}~\cite{graceQ}.

%%%%%%%%%%%%%
\end{acknowledgments}
%%%%%%%%%%%%%

\end{document}